\documentclass[%
 reprint,
superscriptaddress,
 amsmath,amssymb,
 aps,
 prx,
floatfix,
]{revtex4-2}

\usepackage{xcolor}
\usepackage{graphicx}% Include figure files
\usepackage{dcolumn}% Align table columns on decimal point
\usepackage{bm}% bold math
\usepackage{mathtools}
\usepackage{braket}
\usepackage[normalem]{ulem}
\usepackage{subfigure}
\usepackage{appendix}
\usepackage{amsthm}
\newtheorem{theorem}{Theorem}
\newtheorem{conjecture}{Conjecture}

\theoremstyle{definition}

\usepackage{rotating}
\newcommand{\invR}{\begin{sideways}
\begin{sideways}$\mathsf{R}$\end{sideways}\end{sideways}}
\DeclareMathOperator{\arcosh}{arcosh}

\begin{document}

\preprint{APS/123-QED}

\title{Quantum algorithm for approximating the expected value of a random-exist quantified oracle}

\author{Caleb Rotello}
        \affiliation{National Renewable Energy Laboratory, Golden CO, 80401}

\date{\today}% It is always \today, today,

\begin{abstract}
Quantum amplitude amplification and estimation have shown quadratic speedups to unstructured search and estimation tasks. We show that a coherent combination of these quantum algorithms also provides a quadratic speedup to calculating the expectation value of a \textit{random-exist quantified oracle}. In this problem, Nature makes a decision randomly, i.e. chooses a bitstring according to some probability distribution, and a player has a chance to react by finding a complementary bitstring such that an black-box oracle evaluates to $1$ (or True). Our task is to approximate the probability that the player has a valid reaction to Nature's initial decision. We compare the quantum algorithm to the average-case performance of Monte-Carlo integration over brute-force search, which is, under reasonable assumptions, the best performing classical algorithm. We find the performance separation depends on some problem parameters, and show a regime where the canonical quadratic speedup exists.

\end{abstract}

%\keywords{Dynamical mean-field theory}%Use showkeys class option if keyword
\maketitle

%\tableofcontents

\section{\label{sec:intro}Introduction}
Grover's Algorithm~\cite{grover1,grover2}, later generalized to Quantum Amplitude Amplification (QAA) and Quantum Amplitude Estimation (QAE) by Brassard et al.~\cite{Brassard_2002} is a foundational block of quantum algorithms, being one of the earliest quantum algorithms with a provable speedup, and providing the scaffolding for many quantum algorithms and subroutines to this day~\cite{DJORDJEVIC2023491,LU2023100511,WITTEK201441}. QAA can find bitstring(s) that will be accepted by an oracle with quadratically fewer calls to that oracle than a classical computer would require; this problem is NP-Hard, and QAA's quantum speedup for it catalyzed a deeper search into other ways quantum computers can approach this complexity class~\cite{Bennett_1997}.

Positive opinions of QAA waned for a few years due to the ``souffle problem''~\cite{doi:10.1126/science.275.5300.627}, which is the observation that while QAA requires a number of oracle queries proportional to a square root of the inverse fraction of marked elements to find a marked element, one must know this fraction apriori to obtain the square-root speedup. This problem was addressed with an elegant solution by Yoder et al.~\cite{PhysRevLett.113.210501}, which used advances in Quantum Signal Processing (QSP)~\cite{PRXQuantum.2.040203} to create a wavefunction with a guaranteed $1-\delta^2$ overlap with the set of marked states after enough search time.

%\textcolor{red}{this paragraph needs work}
Brassard's other algorithm, QAE, has also become an essential piece of the quantum algorithms catalog. QAE provides a quadratic speedup to the \textit{approximate} counting problem: It estimates the probability of measuring any bitstring that is a True response to the oracle, given an arbitrary superposition of all bitstrings. Montanaro observed this can be used to estimate the expected value of a function quadratically faster than Monte-Carlo integration (MCI)~\cite{Montanaro_2015}. Both QAE and MCI have been shown to be asymptotically optimal~\cite{optimalMCI}: for $N$ samples, MCI samples at the ``Monte Carlo limit'' where error converges as $1/\sqrt{N}$, and QAE samples at the ``Hiesenberg limit'', where error converges as $1/N$. Because of this guaranteed quadratic speedup, QAE is poised to be widely adopted as an essential quantum algorithm for problems
%in the \#P complexity class, 
requiring the estimation of an expectation value of a random variable~\cite{Gacon_2020}, seeing implementations in fields like finance~\cite{Woerner_2019, Stamatopoulos_2020} and power systems~\cite{rotello2024calculatingexpectedvaluefunction}.
Unlike the QAA case, where often the oracle can be ``unpacked'' to find some structure in the problem, thus loosing much if not all of the speedup~\cite{PhysRevX.14.041029}, for QAE there is no oracle to ``unpack'', meaning the speedup survives in practical use cases.

%\textcolor{red}{justify stochastic programs FIX THIS PARAGRAPH}
While deterministic combinatorial problems are contained in (and often complete for) the NP complexity class, tasks involving planning with uncertainty usually lie in the PP or \#P classes~\cite{Littman2001}. Dramatized as games against Nature~\cite{PAPADIMITRIOU1985288}, %\textcolor{red}{more sources}, 
these problems often contain the randomized ($\invR$) and existential ($\exists$) quantifiers on boolean variables, where $\invR x$ means the value of $x$ is decided by some probability distribution (a move made by Nature) and $\exists x$ is a decision variable controlled by the player (these existentially quantified variables behave the same as decision variables in typical NP-complete problems like SAT). The order of these quantifiers is also significant, as $\exists x_1 \invR x_2$ means $x_1$ must be decided \textit{before} the realization of $x_2$. Alternatively, the ordering $\invR x_1 \exists x_2$ means $x_2$ may be decided \textit{after} the uncertainty in $x_1$ is realized as a definite value. 

In this manuscript we solve a random-exist quantified oracle (REQO) stochastic programming task. Given $n$ boolean variables, some probability distribution defined over the first $b$ variables, and an oracle $f$, compute
\begin{equation}
    \invR x_1 \dots \invR x_b \exists y_1 \dots \exists y_{n-b} \big(\mathbb{E}[f(x,y)]),
\end{equation}
where $\mathbb{E}[f]$ is the expected value of the oracle, and Nature gets to move before the player does. In plain English, this asks ``what is the probability that we will have a valid reaction to Nature?'' % \textcolor{red}{Can we just state that it is at x level of counting hierarchy?} %We will argue that this task is $\spnp$-Hard. 
We abbreviate the formula as 
\begin{equation}\label{eq:REQO}
    \invR x \exists y \big(\mathbb{E}[f(x,y)]\big),
\end{equation}
where it is understood that $x = (x_1, \dots x_b)$, $y=(y_1, \dots y_c)$, and $c\equiv n-b$. 
%This specific ordering of quantifiers can be seen for 
Certain stochastic boolean satisfiability formulations, e.g.~\cite{10.5555/3171642.3171741}, can be reduced to Eq.~\ref{eq:REQO} and, if the search problem is swapped out for a min/max problem, it matches previous quantum algorithms work~\cite{rotello2024calculatingexpectedvaluefunction}, as well as the second-stage cost function commonplace in industrial two-stage stochastic programs~\cite{shapiro2021lectures,laporte1992,dempster1981,haneveld2000}.

In the language of QAA, we say that the oracle $f$ has a set of ``marked'' $n$-bit strings. For each bitstring on $x$ that can be chosen by Nature, we will search for a bitstring on $y$ that completes a marked state. Unlike QAA, we are not searching for a marked space; rather, we are finding the probability a string on $x$ is drawn such that it concatenated to a string on $y$ is in the marked space.
Define $\{\xi\} \equiv \Xi \equiv \{0,1\}^b$ as the set of all $b$-bit strings over binary vector $x$, $p(\xi)$ the probability that Nature chooses bitstring $\xi$ from  $\Xi$, and $\{\phi\} \equiv \Phi$ the set of all $c$-bit strings over binary vector $y$. For each $\xi$, there \textit{might} be one or more $\phi$ which satisfies (solicits a True or 1 response from the oracle) such that $f(\xi,\phi)= 1$. We say such a bitstring \textit{completes} $\xi$, because the concatenation of the two bitstrings completes a marked state,
%(to avoid using language from boolean satisfiability), 
denote one such bitstring as $\phi_\xi^*$, and additionally define the set of all such $\phi$ for a specific $\xi$ as $\Phi_\xi^* \equiv \{\phi \; | \; f(\xi,\phi) = 1\}$. Let the set of all $\phi$ which do \textit{not} complete $\xi$ be denoted as $\bar{\Phi}_\xi^*\equiv \Phi\setminus \Phi_\xi^*$. Obviously, if no completing bitstring exists for some $\xi$, then $\Phi_\xi^* =\emptyset$. Note that just because some bitstring $\phi_\xi^*$ satisfies $f$ for bitstring $\xi$ does not imply it satisfies $f$ some other bitstring $\xi'$. We can think of Eq.~\ref{eq:REQO} as computing the probability that, after the realization of $x$, the set $\Phi_\xi^*$ is not empty. Our final piece of notation is $\lambda_\xi\equiv|\Phi_\xi^*| / |\Phi|$, which is the fraction of $c$-bit strings that complete a given $\xi$.

We now outline the remainder of this manuscript. We provide a classical and quantum algorithm to compute an estimate $\tilde \mu$ to the expected value of the REQO, defined as 
\begin{equation}\label{eq:mu}
    \invR x \exists y \Big(\mathbb{E}[f(x,y)] = \mu\Big),
\end{equation}
such that the estimate is within additive $\epsilon$ of the true expected value with a certain probability. We find that, assuming $\lambda=O(1/2^c)$, the quantum algorithm can provide a quadratic speedup, although other regimes for advantage may exist. Importantly, we do \textit{not} claim an advantage holds for all distributions of $\lambda$. 
The classical algorithm, along with its expected oracle query complexity, is described in Sec.~\ref{sec:classical}. The quantum algorithm and its oracle query complexity is described in Sec.~\ref{sec:qalg}; we use canonical quantum amplitude estimation~\cite{Brassard_2002} on oblivious quantum amplitude amplification~\cite{PhysRevLett.113.210501}. Other implementations of these subroutines, like~\cite{Rall_2023, Grinko_2021}, might improve on our results. For completeness, in Sec.~\ref{sec:sharpp} we prove the REQO problem is $\sharp$P-Hard. Finally, we discuss the relative strengths of the algorithm and quantum speedup, as well as address questions posed by this work, in Sec.~\ref{sec:discussion}. %\textcolor{red}{State in terms of standard counting class concentration/approximation bounds. MAIN RESULT HERE}

%\textcolor{red}{address this}Finally, we make the assumptions that all $|\Phi_\xi^*|$ are $O(n)$ (note that this includes zero) and that randomized quantified variables are independently distributed (i.d.). For reasons we will discuss later, our results become less analytically guaranteed as the variance of $\lambda_\xi$ between different $\xi$ increases. 
%Additionally, despite the uniform distribution being sufficient for analytical results of many classical algorithms for the counting complexity classes~\cite{}, arbitrary distributions, particularly ones with high correlation between variables, are exponentially hard to prepare on a quantum computer~\cite{}. This does not change the oracle query complexity of our results, but does have an impact on circuit depth.

%\section{Results}
\section{Classical Algorithm}\label{sec:classical}
This is, under the assumption that no structure of the problem can be leveraged, the best possible classical algorithm to estimate $\mu$. First, we take $N$ samples of $\xi$ from $\Xi$, and order them as $\{\xi_j\}_{j=1}^N$. Then, for each $\xi_j$, we perform a brute-force search for a completing bitstring $\phi^*_{\xi_j}$ such that $f(\xi,\phi^*_{\xi_j})=1$. If a completing bitstring is found, we can terminate the search for $\xi_j$ and move on to $\xi_{j+1}$. If no completing bitstring exists, we will search all possible $c$-bit strings before reaching this conclusion. This will take an expected $1/\lambda_{\xi_j}$ calls to the oracle $f$ in the case where a completing bitstring exists, and $2^c$ otherwise. If $\delta(|\Phi_{\xi_j}^*|)=1$, where $\delta(.)$ is the Kronecker delta, indicates that no completing bitstring exists for some $\xi_j$, then the estimate is given by 
\begin{equation}
    \tilde{\mu} = \frac{1}{N}\sum_{j=1}^N \big(1-\delta(|\Phi_{\xi_j}^*|)\big).
\end{equation}
If we assume that $N<<|\Xi|$, we are unlikely to draw the same $\xi_j$ multiple times. If $\{\lambda_{\xi_j} >0\}$ represents, for the set of $N$ samples, all $\lambda_{\xi_j} \in \{\xi_j\}_{j=1}^N$ which are greater than zero, our query complexity can be written as 
\begin{equation} \tilde\mu\sum_{\lambda\in\{\lambda_{\xi_j>0}\}} \lambda^{-1} + (1-\tilde\mu)2^c.
\end{equation}
According to the central limit theorem, means our \textit{expected} number of queries to $f$ is $(\mu\mathbb{E}_{\lambda_\xi>0}[\lambda_\xi^{-1}]+(1-\mu)2^c)\times N$.

Assuming the variance of $f$ is bound by some $\sigma^2$, the error of this estimate can be bound with Chebyshev's inequality
\begin{equation}
    \text{Pr}\Big[|\mu-\tilde{\mu}| \geq \epsilon\Big] \leq \frac{\sigma^2}{N\epsilon^2}. 
\end{equation}
$f$ is bound on $\{0,1\}$, so by Popoviciu’s inequality on variances we know that $\sigma^2\leq 1/4$~\cite{Popoviciu}. Note that, unless $\sigma^2$ is small as a function of $n$, it only impacts performance as a constant multiple. 
This also implies the classical algorithm requires $N=O(\epsilon^{-2})$ samples from $\Xi$ to reach finitely small success probability. Because we are required by the oracle to perform an unstructured search, no classical algorithm can perform better than $O(\lambda^{-1}_\xi)$ to determine if a completing bitstring exists for a scenario $\xi$. Additionally, MCI is considered asymptotically optimal~\cite{optimalMCI}. %\textcolor{red}{Monte carlo is considered asymptotically optimal for approximating}. 
Therefore, we get the following conjecture.
\begin{conjecture}[Classical approximation algorithm]\label{conj:caa}
    No classical algorithm  produces an estimate $\tilde \mu$ such that $\text{Pr}\big[|\mu - \tilde \mu| \geq \epsilon\big] \leq \sigma^2/(N\epsilon^2)$ with expected oracle query complexity less than 
    $\big(\mu\mathbb{E}_{\lambda_\xi>0}[\lambda_\xi^{-1}]+(1-\mu)2^c\big) N = \big(\mu\mathbb{E}_{\lambda_\xi>0}[\lambda_\xi^{-1}]+(1-\mu)2^c\big)/\epsilon^2$ if the cumulative distribution function of $\lambda$ is unknown.
    %$\mathbb{E}_\xi[\lambda_\xi^{-1}]N=\mathbb{E}_\xi[\lambda_\xi^{-1}]/\epsilon^2$.
\end{conjecture}

The last note about the distribution of $\lambda$ is to cover the following reason: if the cumulative distribution of $\lambda$ is known, for the case where no bitstring completes $\xi$ we can avoid checking all $2^c$ bitstrings. This is because we might have some confidence that no completing bitstring exists once sufficiently many (but not all) have been checked.

\begin{figure}[hpbt]% [hpbt] what you need
     \centering
     \includegraphics[width=\linewidth]{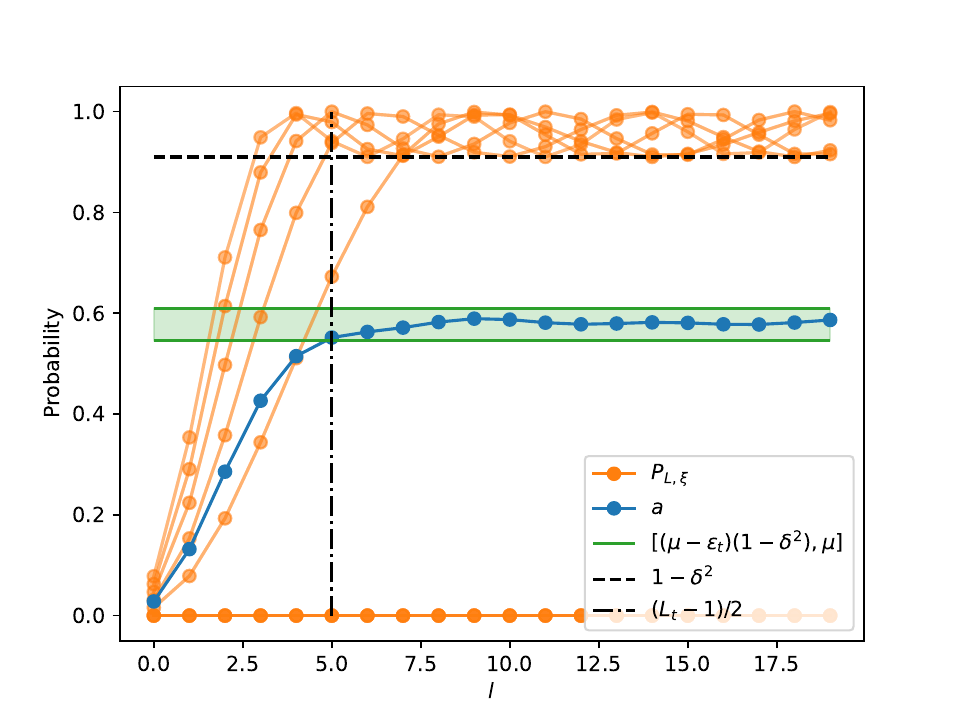}
     \caption{The dynamics of each $\xi$ subspace's success probability $P_{L,\xi}$ as a function of search depth $l$ (defined as $L=2l+1$, described in detail in Sec~\ref{sec:oqaa}). We hold $\delta=.3$ constant; each $P_{L,\xi}$, or the orange lines, follows Eq.~\ref{eq:chebyP}. The horizontal dashed line is at $1-\delta^2$, the convergence floor. The blue line represents the quantity $a$ from Eq.~\ref{eq:a}, should that $L$ be the chosen amplitude amplification search depth; our search routine is converged once $a$ is within the green region. The vertical dashed line represents $L_t$, or the minimum search depth to converge the search, with $\epsilon_t\leq0.01$. Note that not all of the individual searches have converged at $L_t$, but the quantity we will estimate $a$ has. To generate data: $c=b=6$, with an independent identical distribution for the $x$ register. The oracle $f(\xi,\phi)$ returns 1 if $\xi/8 - 3 > \phi$, with $\xi,\phi$ treated as integers, evaluates to true.}\label{fig:lines}
 \end{figure}

\section{Quantum Algorithm}\label{sec:qalg}

The quantum algorithm works by performing quantum amplitude estimation~\cite{Brassard_2002} on a state synthesized by oblivious quantum amplitude amplification~\cite{PhysRevLett.113.210501}. We encode the probability distribution of Nature's possible actions as a quantum wavefunction on $b$ qubits, perform quantum amplitude amplification to search for a valid reaction (completing bitstring) to each scenario at once, and use quantum amplitude estimation to estimate the expectation value of this per-scenario solved wavefunction on the oracle at the Heisenberg limit.

%use a coherent combination of the quantum amplitude amplification and estimation kernels from early quantum search and estimation work. %We describe the amplitude amplification kernel first, and place bounds on the wavefunction it can create.

The quantum amplitude amplification routine, described in detail in Sec.~\ref{sec:oqaa}, uses a unitary operator $\mathcal{S}_L$ and a state preparation operator $V$ to create the \textit{sample-solved wavefunction}
\begin{multline}\label{eq:sswvfn}
    \ket{\mathcal{S}_L} = \mathcal{S}_LV\ket{0}^{\otimes n} = \\
    \sum_{\xi \in \Xi} \sqrt{p(\xi)}\ket{\xi} \Big(\sqrt{1-P_{L,\xi}}\ket{\bar{\Phi}^*_\xi} + \sqrt{P_{L,\xi}}\ket{\Phi_\xi^*}\Big)
\end{multline}
with $L$ calls to the oracle. This routine runs $2^b = |\Xi|$ search problems, or a search for the completing bitstring to each $\xi$, at once. Each search problem has its own success probability $P_{L,\xi}$ defined as 
\begin{equation}\label{eq:chebyP}
    P_{L,\xi} = 1-\delta^2 T_L(T_{1/L}\big(1/\delta)\sqrt{1-\lambda_\xi}\big)^2
\end{equation}
with $T_L$ the $L$th Chebyshev polynomial of the first kind~\cite{chebyshevpoly_1990}
\begin{equation}
    T_L(x) = \begin{cases}
        \cos{(L \arccos{(x)})} &|x| \leq 1\\
        \cosh{(L \arcosh{(x)})} & x \geq 1
    \end{cases}
\end{equation}
($x\leq-1$ is outside the scope of this manuscript). Each success probability is the overlap of the $\ket{xi}$ component of the wavefunction with the set of satisfying solutions for that scenario: $P_{L,\xi} = |\bra{\xi,\Phi^*_\xi}\mathcal{S}_L \ket{\xi,+}|^2$. If, for a given $\xi$, there is no completing bitstring $\phi$ such that $f(\xi,\phi)=1$, then $P_{L,\xi}=0$ for all $L$. Refer to the orange traces in Fig.~\ref{fig:lines} to see the dynamics of $P_{L,\xi}$.

If a scenario $\xi$ is satisfiable, $P_{L,\xi}$ is considered converged for that scenario once it is bound by $1-\delta^2 \leq P_{L,\xi} \leq 1$; this criteria is met when the number of oracle queries $L$ (and, consequently, depth of the $\mathcal{S}_L$ routine) satisfies the inequality 
\begin{equation}
    L \geq \frac{\log(2/\delta)}{\sqrt{\lambda_\xi}}.
\end{equation}
Again, refer to Fig~\ref{fig:lines} to see that, with large enough $L$, each individual search success probability converges to $[1-\delta^2,1]$.
%This presents us with a conundrum - if $L$ is too small, then many of the searches will not converge to the $\delta^2$ window. But if $\lambda_\xi$ has a wide distribution, our quantum algorithm will have to choose $L$ proportional to the smallest $\lambda_\xi$. We can resolve this by picking some $\lambda_t$ apriori such that $\int_{\lambda_t}^1 p(\lambda_\xi) d\xi = 1-\epsilon$. 

This presents a conundrum, as a single $L$ must be chosen such that a large fraction of the individual searches converge, while still minimizing algorithm runtime. To resolve this, we need a way to analyze which searches have converged. This choice of $L$ is made as follows:
Let the probability of drawing a bitstring $\xi$, such that the fraction $\lambda_\xi$ of bitstrings that complete it is equivalent to some target $\lambda$ be given by 
\begin{equation}\label{eq:pr_lambda}
p(\lambda) = \sum_{\xi\in \Xi} p(\xi) \delta(\lambda-\lambda_\xi),
\end{equation}
where $\delta(.)$ is the Kronecker delta function. 
%We then pick some minimum satisfying fraction $1/|\Phi|\leq\lambda_t \leq 1$ threshold such that 
%\begin{equation}\label{eq:epsilont}
%    \int_{\lambda_t}^1 p(\lambda)d\lambda = \mu(1-\epsilon_t),
%\end{equation}
%and define the search depth as $L_t = \Big\lceil \log(2/\delta)/\sqrt{\lambda_t} \,\Big\rceil$. This threshold means that, for our calculation, any scenario $\xi$ such that $\lambda_\xi$ is \textit{beneath} the threshold $\lambda_t$ will not have a search converged to $[1-\delta^2,1]$. Eq.~\ref{eq:epsilont} is structured so that the quantity $\mu(1-\epsilon_t)$ approaches $\mu$ as $\lambda_t$ approaches $1/|\Phi|$; this also reflects the observation that higher $\epsilon_t$ causes less systematic error when $\mu$ is small. Any bitstring $\xi$ such that $\lambda_\xi \geq \lambda_t$ will have a success probability $1-\delta^2\leq P_{L_t,\xi} \leq 1$.
Using this definition, notice that, because there is an integer number of satisfying bitstrings in $\Phi_\xi^*$ for a given $\xi$, there is no $\lambda$ such that $0<\lambda<1/|\Phi|$, and
\begin{equation}\label{eq:alternative_mu}
    \int_{1/|\Phi|}^1 p(\lambda)d\lambda = \mu.
\end{equation}
We pick some minimum satisfying fraction $\lambda_t$ on the interval $1/|\Phi| \leq \lambda_t\leq 1$, and define the oblivious amplitude amplification depth as $L_t = \Big\lceil \log(2/\delta)/\sqrt{\lambda_t} \,\Big\rceil$. In the general case, this $L_t$ may not be large enough to accommodate all $\lambda$; we want to quantify this effect. To do so, break up Eq.~\ref{eq:alternative_mu} as 
\begin{equation}\label{eq:alternative_mu_summed}
    \mu = \int_{1/|\Phi|}^{\lambda_t} p(\lambda)d\lambda + \int_{\lambda_t}^1 p(\lambda)d\lambda.
\end{equation}
In the above formula, the left integral is the probability of drawing a scenario $\xi$ such that $|\Phi_\xi^*|\neq \emptyset$ (a completing bitstring exists for $\xi$), but $\lambda_\xi<\lambda_t$, meaning the search for a satisfying solution has not converged to $[1-\delta^2,1]$ for this scenario. Complementarily, the right integral is the probability of drawing a scenario such that $\lambda_\xi\geq\lambda_t$. The following bi-directional statements can be made:    
\begin{equation}
\begin{split}\label{eq:target_convergence}
    \lambda_\xi \geq \lambda_t \; &\Leftrightarrow \; 1-\delta^2\leq P_{L_t,\xi} \leq 1\\
    \lambda_\xi < \lambda_t \; &\Leftrightarrow \; 0\leq P_{L_t,\xi} < 1-\delta^2.
\end{split}
\end{equation}
It will be useful to re-write Eq.~\ref{eq:alternative_mu_summed} as
\begin{equation}\label{eq:epsilon_t}
    \int_{\lambda_t}^1 p(\lambda)d\lambda = \mu - \int_{1/|\Phi|}^{\lambda_t} p(\lambda)d\lambda = \mu - \epsilon_t,
\end{equation}
to define $\epsilon_t = \int_{1/|\Phi|}^{\lambda_t} p(\lambda)d\lambda $.
%An acceptable $\lambda_t$ quantity is highly dependent on the distribution of $\lambda$. 

Next, quantum amplitude estimation (QAE)~\cite{Brassard_2002}, described in greater detail in Sec.~\ref{sec:qae}, is used to extract an estimate of $\mu$ from the problem statement in Eq.~\ref{eq:mu}. Formally, QAE calculates the probability of receiving a pair $\xi,\phi$ such that $f(\xi,\phi)=1$ from a measurement of the sample-solved wavefunction $\ket{\mathcal{S}_L}$. This is the quantity $a$ such that 
\begin{equation}\label{eq:a}
    a = \sum_{\xi\in\Xi} p(\xi)P_{L_t,\xi}.
\end{equation}
With canonical QAE, for $O(M)$ calls to the state preparation operator $\mathcal{S}_L$, we produce an estimate $\tilde a$ that obeys
\begin{equation}\label{eq:qae_convergence}
    \text{Pr}\Big[|a-\tilde a| \leq 2\pi \frac{\sqrt{a(1-a)}}{M} + \frac{\pi^2}{M^2}\Big] \geq \frac{8}{\pi^2}.
\end{equation}
This process adds $m = \log_2(M)+1$ ancilla qubits. In total, our quantum algorithm uses $(L_t+1)(2M-1) = 2ML_t+2M-L_t-1$ calls to the oracle $f$ (for the complete derivation of this, see Sec~\ref{sec:qae}).

Lastly, we characterize the error of the algorithm. First, we bound the distance between $a$ and $\mu$. Recall that we have chosen some $L_t$ which can be used to predict these bounds. Combining Eq.~\ref{eq:a} with Eq.~\ref{eq:target_convergence}, we can see the upper bound on $a$ is $a\leq \mu$. The lower bound is a little more involved; first, partition $\Xi$ into two subsets $\Xi_{\geq t}=\{ \xi \, | \, \lambda_\xi \geq \lambda_t\}$ and $\Xi_{<t}=\{\xi \, |\, \lambda_\xi < \lambda_t\}$. Obviously, $\Xi = \Xi_{\geq t} \cup \Xi_{< t}$. Using Eq.~\ref{eq:a} again, we can see
\begin{align}
    a &= \sum_{\xi\in\Xi_{\geq t}} p(\xi)P_{L_t,\xi} + \sum_{\xi\in\Xi_{< t}}p(\xi)P_{L_t,\xi}.
\end{align}
Notice the rightmost sum is $\geq 0$, and plugging in the observation from Eq.~\ref{eq:target_convergence} for $\lambda_\xi \geq \lambda_t$, 
\begin{align}
    a&\geq \sum_{\xi\in\Xi_{\geq t}} p(\xi)P_{L_t,\xi} \geq \big(1-\delta^2\big)\sum_{\xi\in\Xi_{\geq t}} p(\xi).
\end{align}
Using the definition of $\Xi_{\geq t}$ and Eq.~\ref{eq:pr_lambda}, we can see that $\int_{\lambda_t}^1p(\lambda)d\lambda=\sum_{\xi\in\Xi_t}p(\xi)$; combined with Eq.~\ref{eq:epsilon_t}, we get $\sum_{\xi\in\Xi_{\geq t}} p(\xi) = \mu-\epsilon_t$. Plugging this in to the above, see 
\begin{align}
    a&\geq (\mu-\epsilon_t)(1-\delta^2).
\end{align}
This means $(\mu-\epsilon_t)(1-\delta^2) \leq a \leq \mu$. This is the green window shown in Fig~\ref{fig:lines}, which $a$ converges to. Plugging in Eq.~\ref{eq:qae_convergence} and removing the absolute value, we can see that with probability $\geq 8/\pi^2$
\begin{multline}
    a - 2\pi \frac{\sqrt{a(1-a)}}{M} - \frac{\pi^2}{M^2} \leq \tilde{a} \\\leq \mu +2\pi \frac{\sqrt{a(1-a)}}{M} + \frac{\pi^2}{M^2},
\end{multline}
which can be further simplified by replacing $a$ and using the bound $\sqrt{a(1-a)}\leq 1/2$:  
\begin{equation}\label{eq:q_error}
    (\mu-\epsilon_t)(1-\delta^2) - \frac{\pi}{M} - \frac{\pi^2}{M^2} \leq \tilde{a} \\ \leq \mu+ \frac{\pi}{M} + \frac{\pi^2}{M^2}.
\end{equation}
Finally, we can re-write this as 
\begin{multline}
    \mu-\epsilon_t -\mu\delta^2 +\epsilon_t \delta^2 - \frac{\pi}{M} - \frac{\pi^2}{M^2} \leq \tilde{a} \\ \leq \mu + \frac{\pi}{M} + \frac{\pi^2}{M^2}.
\end{multline}
This can be summarized by, and proves, the following theorem.
\begin{theorem}[Oracle query complexity]\label{thm:oqc}
    A quantum algorithm exists which can compute an estimate $\tilde{a}$ such that $\text{Pr}\big[|\tilde{a}-\mu| \leq \epsilon\big] \geq 8/\pi^2$, where $\epsilon\leq \epsilon_t + \delta^2\mu - \delta^2\epsilon_t +\frac{\pi}{M} + \frac{\pi^2}{M^2}$, with $(L_t+1)(2M-1)$ calls to the oracle $f$.
    % and $M$ calls to the probability distribution preparation $\mathcal{P}$.
\end{theorem}

Note that the bounds on this theorem are not tight; because the algorithm is more likely to underestimate $\mu$ than it is to overestimate it, we bound $\epsilon$ with the underestimate.
%In total, our quantum algorithm uses $(L+1)(M-1) = 2ML+2M-L-1$ calls to the oracle $f$, to produce an estimate $\tilde a$ such that $|\mu-\tilde{a}|\leq\epsilon$ with probability $\geq8/\pi^2$, where $\epsilon\leq \epsilon_t + \delta^2 - \epsilon_t\delta^2 +\frac{\pi}{M} + \frac{\pi^2}{M^2}$. We have set $\mu=1$ for this bound to make it asymptotic.

%$\tilde\mu$, of $\mu$ from the problem statement in Eq.~\ref{eq:mu}

\begin{figure*}[hpbt]% [hpbt] what you need
     \centering
     \includegraphics[width=\linewidth]{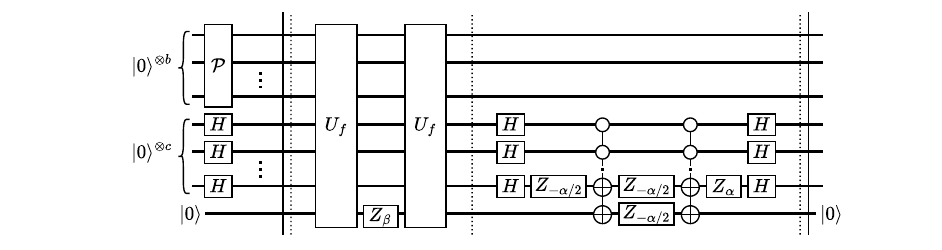}
     \caption{The circuit diagram for the algorithm $S_LV$. The first $b$ qubits in the top register host the probability distribution, or randomly-quantified variables, while the last $c$ qubits host the decision or existentially-quantified variables. As we can see, the oracle unitary $U_f$ is operationally identical to the oracle in canonical quantum amplitude amplification algorithms. The mixing unitary only needs to operate on the decision register.}\label{fig:circuit_oaa}
 \end{figure*}

\subsection{Oblivious Quantum Amplitude Amplification over a Probability Distribution}\label{sec:oqaa}
Here we describe the subroutine used to prepare the wavefunction in Eq.~\ref{eq:sswvfn}; see Theorem~\ref{thm:oaa} for a proof of correctness. This process follows the prescription of~\cite{PhysRevLett.113.210501}, with our own additions of an auxiliary register to store the probability distribution and control the amplitude amplification phase oracle. We set the first $b$ qubits aside for the randomly quantified variables $x$, and the next $c=n-b$ qubits for the existentially quantified variables $y$. First we define the initial state of the algorithm, which is a probability distribution in the $x$ register and a uniform superposition in the $y$ register. We then describe how a quantum amplitude amplification creates the sample-solved wavefunction introduced in Eq.~\ref{eq:sswvfn}, which is the probability distribution on the $x$ variables, with each $\xi$ paired to its completing bitstring $\phi_\xi^*$, should it exist.

Given that $p(\xi)$ is the probability that bitstring $\xi$ is drawn, define the unitary channel $\mathcal{P}$ as
\begin{equation}
    \mathcal{P}\ket{0}^{\otimes b} = \sum_{\xi\in\Xi}\sqrt{p(\xi)}\ket{\xi},
\end{equation}
which encodes the probability distribution over $x$ as a quantum wavefunction with probability $p(\xi)$ of measuring state $\ket{\xi}$. For some probability distributions $\mathcal{P}$ has $O$(poly($b$)) circuit depth, e.g. normal, log-normal, independent, or otherwise learnable distributions~\cite{grover2002creating, Plesch_2011, dasgupta2022loading, Zoufal_2019}, while it requires exponential circuit depth in the general case. %As such, our results are given in terms of ``number of calls to $\mathcal{P}$''. 
We prepare the $y$ register in an even superposition, completing the description for the operator $V$ from Eq.~\ref{eq:sswvfn} as
\begin{equation}
    V= \Big(\mathcal{P}\otimes \mathbb{I}_{2^c}\Big) \times \Big(\mathbb{I}_{2^b} \otimes H^{\otimes c}\Big).
\end{equation}

It will be useful to define the Grover iterate, which is comprised of a reflection $S_{\Phi}(\beta)$ about the target state and a reflection $S_+(\alpha)$ about the initial state, both parameterized by an angle through quantum signal processing~\cite{PRXQuantum.2.040203}. Because we only apply the reflection about the initial state to the $y$ register, our case differs from common implementations of Grover-type algorithms. Each possible scenario $\xi\in\Xi$ can be thought of as a label for its own Bloch sphere, with the $S_t$ and $S_s$ reflections being transformations in SU($2$) on each scenario. This is why our search algorithm's query depth depends on the individual satisfying fractions, $\lambda_\xi$.%, and not the actual number of possible scenarios.

We have access to the oracle $f$, where $f(\xi,\phi)=1$ is a satisfying statement to the oracle and $\phi$ is a completing bitstring for scenario $\xi$. We use this to define a unitary oracle operator 
\begin{equation}\label{eq:oracle_unitary}
    U_f\ket{\xi,\phi,z} = \ket{\xi,\phi,z\oplus f(\xi,\phi)},
\end{equation}
which can, in turn, be used to create the reflection about the target states (see Figure~\ref{fig:circuit_oaa}):
\begin{multline}\label{eq:s_phi}
    S_{\Phi} (\beta) = U_f Z_{n+1}(\beta) U_f\\
    =\mathbb{I}_{2^n} - (1-e^{-i\beta})\sum_{\xi\in\Xi} \Big(\ket{\xi}\bra{\xi} \otimes \sum_{\phi \in \Phi_\xi^*}\ket{\phi} \bra{\phi}\Big).
\end{multline}
Note that the target state is labeled by the bitstring $\xi$, while the oracle still need only be applied twice.

We also create the reflection about the initial state. We must clarify this is a reflection about the initial state in the $y$, or decision variable, register only. If the mixing unitary was also applied to the distribution register (first $b$ qubits), we would end up amplifying \textit{all complete bitstrings}, instead of amplifying the completing bitstring, should it exist, for each scenario. In other words, the wavefunction stored in the $x$ register is untouched by this operator. See Figure~\ref{fig:circuit_oaa} for a circuit description. This operator acts as 
\begin{equation}\label{eq:s_plus}
    S_+(\alpha) = \mathbb{I}_{2^n} - (1-e^{-i\alpha})\mathbb{I}_{2^b}\otimes \ket{+}\bra{+}^{\otimes c}. 
\end{equation}
%It will be useful to re-write it as 
%\begin{multline}
%    S_+(\alpha)= \mathbb{I}_{2^n} - (1-e^{-i\alpha})
%    \sum_{\xi\in\Xi} \ket{\xi}\bra{\xi}\otimes \\\Big((1-\lambda_\xi)\ket{\bar{\Phi}^*_\xi}\bra{\bar{\Phi}^*_\xi} 
%    \\+ \sqrt{\lambda_\xi(1-\lambda_\xi)} \ket{\bar{\Phi}^*_\xi}\bra{\Phi_\xi^*} + \sqrt{\lambda_\xi(1-\lambda_\xi)} \ket{\Phi^*_\xi}\bra{\bar{\Phi}^*_\xi} 
%    \\+ \lambda_\xi\ket{\Phi^*_\xi}\bra{\Phi^*_\xi}\Big),
%\end{multline}
%using $\mathbb{I}_{2^b} = \sum_{\xi\in\Xi} \ket{\xi}\bra{\xi}$.

The Grover iterate is defined as $G(\alpha,\beta) = -S_+(\alpha)S_{\Phi}(\beta)$, and a sequence of $l$ such iterates is our oblivious quantum amplitude amplification algorithm:
\begin{equation}\label{eq:oqaa_alg}
    \mathcal{S}_L = G(\alpha_l,\beta_l)\dots G(\alpha_1,\beta_1) = \prod_{j=1}^l G(\alpha_j,\beta_j),
\end{equation}
which has oracle query complexity $L-1= 2l$. The phases, $\alpha$ and $\beta$, are analytically chosen according to their index and the depth $l$ of $\mathcal{S}_L$:
\begin{multline}\label{eq:alphabeta}
    \alpha_j = -\beta_{l-j+1} = 2\cot^{-1}\Big(\\ \tan{\big(2\pi j/L\big)}\sqrt{1-1/T_{1/L}(1/\delta)^2}\,\Big).
\end{multline}
%where $T_L(x)$ is the Chebyshev polynomial defined as 
This sets up the following
\begin{theorem}[Oblivious amplitude amplification over a probability distribution]\label{thm:oaa}
    The state $\ket{\mathcal{S}_L}$ can be prepared as $\mathcal{S}_LV\ket{0}^{\otimes n} = \ket{\mathcal{S}_L}$, where unitary operator $\mathcal{S}_L$ uses $2l+1$ queries to the boolean oracle $f$.% and $V$ is one query to the probability distribution $\mathcal{P}$.
    %for $l$ calls to the Grover iterate $G(\alpha,\beta)=-S_+(\alpha)S_\Phi(\beta)$
\end{theorem}

\textit{Proof:}
To see that $\mathcal{S}_LV\ket{0}^{\otimes n} = \ket{\mathcal{S}_L}$, we show that a Grover iterate $G(\alpha,\beta)$ acts on each SU(2) subspace labeled by its scenario $\xi$ as it would in the original oblivious amplitude amplification work by Yoder et al.~\cite{PhysRevLett.113.210501}. 
According to~\cite{PhysRevLett.113.210501}, given the operators 
\begin{align}
    \label{eq:U_phi}
    U_\Phi(\beta) &= \mathbb{I}_{2^n} - (1-e^{-i\beta})\sum_{\phi\in\Phi^*} \ket{\phi}\bra{\phi},\\
    \label{eq:U_plus}
    U_+(\alpha) &= \mathbb{I}_{2^n} - (1-e^{-i\alpha}) \ket{+}\bra{+}^{\otimes n}
\end{align}
the following, if $\alpha,\beta$ are chosen with Eq.~\ref{eq:alphabeta}, is true:
\begin{multline}
\prod_{j=1}^l \Big[-U_+(\alpha_j)U_\Phi(\beta_j) \Big] \ket{+}^{\otimes n} \\= \sqrt{1-P_L}\ket{\bar{\Phi}^*} + \sqrt{P_L}\ket{\Phi^*}.
\end{multline}

The operators $U_\Phi$ and $U_+$ act on a SU(2) space. As such, we can show that $S_\Phi$ and $S_+$ act on each SU(2) subspace, as labeled by $\xi$, the same way that $U_\Phi$ and $U_+$ act on a single SU(2) space. 

%The operator $S_\Phi$ in Eq.~\ref{eq:s_phi} is already in this form. Notice that Eq.~\ref{eq:s_phi} can also be written as 
%\begin{equation}
%    S_\Phi = \mathbb{I}_{2^n} - (1-e^{-i\beta})\sum_{\xi\in\Xi} \ket{\xi}\bra{\xi}\otimes \Big( -U_{\Phi_\xi^*}(\beta) + I_{2^n} \Big)
%\end{equation}
 
The operator $S_\Phi$ in Eq.~\ref{eq:s_phi} is already in this form. Notice that, because of the projector $\ket{\xi}\bra{\xi}\otimes(\sum_{\phi\in\Phi_\xi^*}\ket{\phi}\bra{\phi})$, the phase $-(1-e^{-i\beta})$ is only applied to a state $\ket{\xi}\ket{\phi}$ if $f(\xi,\phi)=1$. This means we apply the phase to a bitstring $\phi$ iff $\phi$ is satisfying for a given $\xi$, meaning we have correctly built the phase unitary from Eq.~\ref{eq:U_phi} for each SU(2) subspace.

To see this is also true for $S_+$, we use the fact that $\Xi=\{0,1\}^b$ and thus forms a complete basis. This means $\mathbb{I}_{2^b} = \sum_{\xi\in\Xi} \ket{\xi}\bra{\xi}$; we can replace the identity in Eq.~\ref{eq:s_plus}, giving us 
\begin{equation}
    S_+ = \mathbb{I}_{2^n} - (1-e^{-i\alpha})\sum_{\xi \in \Xi} \ket{\xi}\bra{\xi}\otimes\ket{+}\bra{+}^{\otimes c},
\end{equation}
which will function the same as $U_+$ in Eq.~\ref{eq:U_plus} with a label for each $\xi$. Note that this still holds when $p(\xi)=0$, as the sample $\xi$ is still a basis vector, it just has no chance of being chosen as Nature's move.

%If the mixing unitary was also applied to the distribution register (first $b$ qubits), we would end up amplifying the \textit{scenarios that have a satisfying bitstring}, instead of amplifying the satisfying bitstring, should it exist, for each scenario.
%$\gamma^{-1} = T_{1/L}(1/\delta)$

\subsection{Quantum amplitude estimation}\label{sec:qae}
We will describe the quantum amplitude estimation (QAE) primitive here. This subroutine functions identically to the canonical case; as such, we only describe it, and do not provide any novel observations. From Sec.~\ref{sec:oqaa} we know that 
%\begin{equation}
\begin{multline}
    \mathcal{S}_LV\ket{0}^{\otimes n} = \sum_{\xi\in\Xi}\sqrt{p(\xi)}\Big(\sqrt{1-P_{L,\xi}}\ket{\bar\Phi_\xi^*} \\+ \sqrt{P_{L,\xi}}\ket{\Phi_\xi^*}\Big).
\end{multline}
%\end{equation}
Now add an ancilla qubit initialized as $\ket{0}$ and operate with $U_f$ from Eq.~\ref{eq:oracle_unitary} to separate the wavefunction into a linear combination of a ``good'' state $\ket{\psi_1}$ marked with a $\ket{1}$ in the ancilla and a ``bad'' state $\ket{\psi_0}$ marked with a $\ket{0}$ in the ancilla:
\begin{multline}
    U_f\ket{\mathcal{S}_L}\ket{0} = \\
    \sum_{\xi\in\Xi}\sqrt{p(\xi)}\Big(\sqrt{1-P_{L,\xi}}\ket{\bar\Phi_\xi^*}\ket{0} + \sqrt{P_{L,\xi}}\ket{\Phi_\xi^*}\ket{1}\Big)\\
    = \sqrt{1-a}\ket{\psi_0}\ket{0} + \sqrt{a}\ket{\psi_1}\ket{1}.
\end{multline}
If we were to measure the ancilla qubit, we would receive $\ket{1}$ with probability 
$a$, which can be expressed as the $\mu+O(\epsilon)$ quantity from Eq.~\ref{eq:a},
\begin{equation}
    a = \sum_{\xi\in\Xi}p(\xi)P_{L,\xi}.
\end{equation}
It will be uesful to define the operator $\mathcal{A}=U_f\mathcal{S}_LV$ such that
\begin{equation}\label{eq:init_A}
    \mathcal{A}\ket{0}^{\otimes n}\ket{0} = \sqrt{1-a}\ket{\psi_0}\ket{0} + \sqrt{a} \ket{\psi_1}\ket{1}
\end{equation}

\begin{figure}[]% [hpbt] what you need
\hspace*{-1.5cm} 
     \includegraphics[width=1\linewidth]{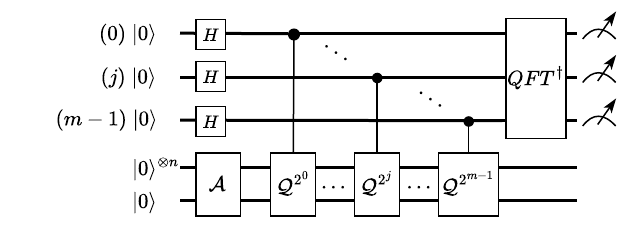}
     \caption{The circuit diagram for QAE.}\label{fig:circuit_qae}
 \end{figure}

To estimate $a$, QAE applies quantum phase estimation to the operator
\begin{multline}
    \mathcal{Q} = \mathcal{A}\big(\mathbb{I}_{2^{n+1}}-2\ket{0}\bra{0}^{\otimes n+1}\big)\mathcal{A}^{\dagger}\\\big(\mathbb{I}_{2^{n+1}}-2\ket{\psi_0}\ket{0}\bra{\psi_0}\bra{0}\big)\\
    = \mathcal{A}\big(\mathbb{I}_{2^{n+1}}-2\ket{0}\bra{0}^{\otimes n+1}\big)\mathcal{A}^{\dagger}\big(\mathbb{I}_{2^n}\otimes-Z_{n+1}\big).
\end{multline}
If $\mathcal{A} = U_f\mathcal{S}_LV$ then $\mathcal{A}^\dagger = V^\dagger\mathcal{S}_L^\dagger U_f$. This is possible for known circuits by reversing the circuit order and negating the angles; it is more complicated but still possible if $\mathcal{P}$ is not known~\cite{Yoshida_2023}.
We add $m$ ``sample'' qubits initialized in $\ket{+}^{\otimes m}$, labeled $j\in[0,\dots,m-1]$. Recall that $M=2^m$. For each sample qubit, we perform $\ket{1}\bra{1}_j\otimes\mathcal{Q}^{2^j}$. Finally, we perform an inverse quantum fourier transform, $QFT^\dagger$, on the sample qubits, and measuring them in the $Z$ basis. This circuit can be seen in Fig.~\ref{fig:circuit_qae}.

The final measurement produces some $m$-bit string $d$; interpreted as an integer, we calculate our estimate as
\begin{equation}
    \tilde a = \sin^2(d\pi / M).
\end{equation}
This estimate obeys the inequality
\begin{equation}
    \text{Pr}\Big[|\tilde a-a|\leq2\pi\frac{\sqrt{a(1-a)}}{M}+\frac{\pi^2}{M^2}\Big] \geq \frac{8}{\pi^2}.
\end{equation}

The estimation routine applies the operator $\mathcal{Q}$ a total of $\sum_{j=0}^{m-1} 2^j$ times, which simplifies to $2^m-1=M-1$ times. As each application of $\mathcal{Q}$ runs $\mathcal{A}$ and $\mathcal{A}^\dagger$, the operator $\mathcal{A}$ or its Hermitian conjugate is run $2(M-1)+1=2M-1$ times total. From Sec.~\ref{sec:oqaa} and Thm.~\ref{thm:oaa} each application of $\mathcal{A}$ or its inverse has $L+1$ calls to the oracle. This echoes our result in Sec.~\ref{sec:qalg}: the total number of oracle calls is $(2M-1)(L+1)$. 

\section{Computational complexity of estimating the expectation value}\label{sec:sharpp}
To complete our argument, we also prove that calculating $\mu$ such that $\invR x\exists y (\mathbb{E}[f(x,y)] = \mu)$ is $\sharp$P-Hard. This can be shown by reduction from graph reliability:
%\begin{definition}[Graph reliability]
Given a graph $\mathcal{G}=(\mathcal{V},\mathcal{E})$ of vertices $\mathcal{V}$ and edges $\mathcal{E}$, compute the reliability of the network, given as the probability that two fixed nodes $u$ and $v$ are connected, when each edge has a $1/2$ probability of failing. 
%\end{definition}
This problem is $\sharp$P-Complete~\cite{dyer2005,Valiant1979TheCO}. Note that we do not claim our algorithm provides a speedup to this problem; this problem is only introduced for the purposes of a proof.

\begin{theorem}[Complexity of computing the expected value of the random-exist quantified oracle]
    It is $\sharp$P-Hard to compute $\mu$ such that $\invR x\exists y \big(\mathbb{E}[f(x,y)] = \mu\big)$.
\end{theorem}

\textit{Proof:} Suppose we are given an instance of graph reliability on a graph $\mathcal{G}=(\mathcal{V},\mathcal{E})$ with target nodes $u$ and $v$. Set $b=c=|\mathcal{E}|$ qubits for the $x$ and $y$ (Nature and player) registers. Let the state $0$ on bit $j\in[0,b)$ (register $x$) imply the $j$th edge has failed, and let the state $1$ on bit $j\in[b,c)$ (register $y$) imply that edge $j$ is included in the path.

An oracle $g$ takes a $b$-bit (or $c$-bit) string $z$, which represents a path, as an input, and tells us if $u$ and $v$ are connected by edges in $\mathcal{E}$ by that string $z$; $g(z)=1$ implies that one can walk between $u$ and $v$ following instructions from string $z$. Thus, for an attempted path $\phi$ in the $y$ register and scenario $\xi$, the oracle on $g(\xi \land \phi)$ will be $0$ if the failures in the specific scenario prohibit traversal, and will be $1$ if there is still a valid path.

Therefore, if we define $f(x,y) = g(x\land y)$, we can see that the reliability of the graph equals $\mu$ such that $\invR x \exists y \big(\mathbb{E}[f(x,y)]=\mu\big)$.

\section{Discussion of advantage}\label{sec:discussion}
We describe some situations where the quantum algorithm has lower query complexity to reach the same additive error $\epsilon$. 
From Sec.~\ref{sec:classical} Conj.~\ref{conj:caa} and Sec.~\ref{sec:qalg} Thm.~\ref{thm:oqc}, the classical query complexity is 
\begin{equation}
    \big(\mu\mathbb{E}_{\lambda_\xi>0}[\lambda_\xi^{-1}] +(1-\mu)2^c\big)/\epsilon^2
\end{equation}
and the quantum query complexity is
\begin{equation}
    (L_t+1)(2M-1).
\end{equation}
For the quantum algorithm, we have $\epsilon\leq \epsilon_t+\delta^2\mu-\delta^2\epsilon_t +\frac{\pi}{M} + \frac{\pi^2}{M^2}$. As stated in the Introduction, and visible here, the performance is mostly characterized by the distribution of the parameter $\lambda$. There is a quantum advantage in oracle query complexity whenever the following holds:
\begin{equation}
    (L_t+1)(2M-1) < \big(\mu\mathbb{E}_{\lambda_\xi>0}[\lambda_\xi^{-1}] +(1-\mu)2^c\big)/\epsilon^2.
\end{equation}

Assume that $\lambda=O(1/2^c)$. Recall that $L_t = \lceil\log_2(2/\delta)\rceil/\sqrt{\lambda_t}$. If we substitute for $\lambda$, the total oracle complexity is $(\lceil\log_2(2/\delta)/\sqrt{2^c}\rceil +1)(2M-1)$. In this case, $\epsilon_t=0$, so $\delta+\pi/M+\pi^2/M^2 = \epsilon$ is sufficient. If we let $\delta=\epsilon/2$ and $M=2\pi/\big(\sqrt{2\epsilon+1}-1\big)$ this criteria is met; we can re-write the quantum query complexity as 
\begin{equation}
    (\Big\lceil\sqrt{2^c}\log_2(4/\epsilon)\Big\rceil +1)(\frac{4\pi}{\big(\sqrt{2\epsilon+1}-1\big)}-1).
\end{equation}
In Big-O notation (dropping the logarithmic factor for simplicity) we get $\tilde O\big(\sqrt{2^c}/\epsilon\big)$.
If we re-write the classical query complexity as $O\big(2^c/\epsilon^2\big)$, we can see that, when $\lambda=O(1/2^c)$, the quantum algorithm has a quadratic advantage.

An open question is how the query complexity might separate for other distributions of $\lambda$, for known and unknown cumulative distribution functions (CDF) of $\lambda$. If the CDF is unknown, for a single $\xi$ the classical algorithm needs to check all $2^c$ $c-$bit strings if no completing one exists. However, the quantum algorithm only needs to check up to some $\lambda_t>0$, instead of all $2^c$ strings, to reach convergence. If the quantum algorithm can try a few $\lambda_t$, this might open the door for exponential speedups at $\lambda=O(1)$. The existence of such a performance separation is an open question. If the CDF is known, it is likely a classical algorithm exists where we can \textit{assume} no completing string exists after enough searching; it is an open question how the performance of each algorithm can be bound in this case.

%For example, consider if $\lambda$ were large fractions, but the distribution unknown. 
%there quantum algorithm could terminate with only a few calls to the search routine. However, the classical algorithm still needs to search the entire space to confirm a negative

\section{Conclusion}
In this work, we discussed a classical and quantum algorithm for a stochastic programming task: compute the probability we have a valid reaction to an event caused by Nature. This problem, for an oracle function, was shown to be $\sharp$P-Hard. The classical algorithm works by drawing random samples and performing brute-force search on each sample, while the quantum algorithm encodes Nature's potential moves as a quantum wavefunction, solves the search problem over search scenario (basis vector) in the wavefunction via oblivious amplitude estimation, then extract an estimate via quantum amplitude estimation. The performance separation of the algorithms depends on the fraction of bitstrings that are completing bitstrings for a given scenario, as well as the distribution of this parameter across different scenarios. We show that the canonical quadratic separation seen in quantum amplitude estimation and amplification exists when $\lambda=O(1/2^c)$. This work fits into a broader effort to use quantum computers to solve problems with stochasticity (see e.g.~\cite{rotello2024calculatingexpectedvaluefunction,Woerner_2019,Gacon_2020}) by showing how a quantum computer can solve many problems at once, with each problem encoded by some basis vector in a wavefunction, and how to estimate statistics of the solved problems at the Heisenberg limit.

\begin{acknowledgments} 
%This work was authored in part by the National Renewable Energy Laboratory, managed and operated by the Alliance for Sustainable Energy, LLC for the U.S. Department of Energy (DOE) under Contract No. DE-AC36-08G028308. A portion of this research was performed using computational resources sponsored by the
%U.S. Department of Energy’s Office of Energy Efficiency and Renewable Energy and located at the National Renewable Energy Laboratory.
The views expressed in the article do not necessarily represent the views of the DOE or the U.S. Government. The U.S. Government retains and the publisher, by accepting the article for publication, acknowledges that the U.S. Government retains a nonexclusive, paid-up, irrevocable, worldwide license to publish or reproduce the published form of this work, or allow others to do so for U.S. Government purposes.
\end{acknowledgments}

\bibliography{bibliography}% Produces the bibliography via BibTeX.
%\printbibliography
%\include{supplementary_information}

\end{document}